\RequirePackage{amsthm}
\pdfoutput=1
\documentclass[pdflatex,sn-mathphys-num]{sn-jnl}% Math and Physical Sciences Numbered Reference Style 
\usepackage{lmodern}
\usepackage{graphicx}%
\usepackage{multirow}%
\usepackage{amsmath,amssymb,amsfonts}%
\usepackage{amsthm}%
\usepackage{mathrsfs}%
\usepackage[title]{appendix}%
\usepackage{xcolor}%
\usepackage{textcomp}%
\usepackage{manyfoot}%
\usepackage{booktabs}%
\usepackage{algorithm}%
\usepackage{algorithmicx}%
\usepackage{algpseudocode}%
\usepackage{listings}%
\usepackage{upgreek}
\theoremstyle{thmstyleone}%
\theoremstyle{thmstyletwo}%

\theoremstyle{thmstylethree}%

\begin{document}

\title[Article Title]{Topology optimization of high-performance optomechanical resonator}

\author*[1]{\fnm{Yincheng} \sur{Shi}}\email{yishi@dtu.dk}

\author[2]{\fnm{Fengwen} \sur{Wang}}

\author[1]{\fnm{Dennis} \sur{Høj}}

\author[2]{\fnm{Ole} \sur{Sigmund}}

\author[1]{\fnm{Ulrik} \sur{Lund Andersen}}

\affil[1]{\orgdiv{Center for Macroscopic Quantum States (bigQ), Department of Physics}, \orgname{Technical University of Denmark}, \orgaddress{\street{Fysikvej}, \city{Kgs. Lyngby}, \postcode{2800},  \country{Denmark}}}

\affil[2]{\orgdiv{Department of Civil Mechanical Engineering}, \orgname{Technical University of Denmark}, \orgaddress{\street{Koppel's Allé 404}, \city{Kgs. Lyngby}, \postcode{2800}, \country{Denmark}}}

%%==================================%%
%% Sample for unstructured abstract %%
%%==================================%%

\abstract{
High quality mechanical resonators are critical for driving advances in quantum information technologies, precision sensing, and optomechanics. However, achieving compact resonator designs that maintain high performance is a key challenge. In this study, we present a new class of compact resonators optimized to operate at higher-order eigenmodes, achieving both high frequencies and enhanced quality factor-frequency (\textit{Qf}) products. By employing topology optimization to maximize the damping dilution factor, these resonators achieve minimized edge bending losses and enhanced intrinsic damping. Their high-(\textit{Qf}) performance and compact form factor position these resonators as promising candidates for applications in quantum information transduction, advanced optomechanical systems, and next-generation sensing technologies.
%This study proposes novel resonators designed to operate at higher-order eigenmodes, achieving both high frequencies and quality factor-frequency (\textit{Qf}) product in a compact form factor. These designs are optimized by maximizing the damping dilution factor via topology optimization. Ringdown measurements  show that over 75\% of the fabricated samples reach the quantum regime at room temperature. The high \textit{Qf} products of these resonators significantly improve force sensitivity and satisfy ground-state cooling requirements, highlighting their potential for applications in precision sensing, detection, and advanced quantum technologies.
}


%%\pacs[JEL Classification]{D8, H51}

%%\pacs[MSC Classification]{35A01, 65L10, 65L12, 65L20, 65L70}

\maketitle

\section{Introduction}\label{sec1}
State-of-the-art mechanical resonators with high quality factors (\textit{Q}) and high quality-factor-frequency (\textit{Qf}) products are essential for applications ranging from precision force and acceleration sensing \cite{bib32, bib33} to magnetic spin detection \cite{bib34} and advanced quantum information technologies. These technologies include the development of quantum transducers for converting superconducting microwave qubits or spin qubits to optical qubits \cite{bib35}, as well as applications in quantum memory and phononic computing.

Ultra-high quality factors—exceeding 100,000 times the intrinsic material limits of the mechanics \cite{bib2, bib39}—can be achieved through dissipation dilution, a phenomenon first observed in gravitational-wave detectors \cite{bib37} and later in Si$_3$N$_4$ nano-strings \cite{bib38}. Dissipation dilution is achieved through tensile strain and geometric optimization, which allows these structures to store elastic energy in tension or through geometric deformations, significantly reducing bending losses and enhancing \textit{Q} \cite{bib45, bib2}. This effect is quantified by the dilution factor \textit{D}$_q = Q/Q_0$, where $Q_0$ represents the intrinsic quality factor \cite{bib40, bib2}

%State-of-art mechanical resonator with high quality factor (\textit{Q}) and high quality-factor-frequency (\textit{Qf}) product are crucial for applications in force \cite{bib32}, acceleration sensing \cite{bib33}, magnetic spin detection \cite{bib34} and various quantum technologies. These include the optical readout of superconducting qubits via microwave-to-optical conversion \cite{bib35} and ground state cooling at room temperature using a membrane-in-the-middle scheme \cite{bib36}. Such resonators can achieve quality factors, \textit{Q}, that exceed by more than 100 thousand times the values predicted by intrinsic material losses,  a phenomenon known as damping dilution \cite{bib2, bib39}. This effect was first observed in gravitational-wave detectors\cite{bib37} and later in a Si$_3$N$_4$ nano-string \cite{bib38}. It is quantified by the dilution factor \textit{D}$_q = Q/Q_0$, where $Q_0$ represents the intrinsic quality factor \cite{bib40, bib2}. It is generally understood that the elastic energy stored in the tension of strings or membranes, or through geometrically nonlinear deformation \cite{bib39}, is effectively lossless, leading to an increase in \textit{Q} \cite{bib45, bib2}. 

A basic understanding of damping dilution can be gained through models of pre-stressed strings and membranes \cite{bib2}. These models show that the dilution factor \textit{D}$_q$ is positively correlated with the aspect ratio (length-to-thickness) and strain, achievable through nanofabrication techniques like LPCVD deposition of pre-stressed Si$_3$N$_4$ thin films. Additionally, \textit{D}$_q$ is inversely proportional to edge bending terms and inversely quadratic with respect to anti-nodal bending terms. By minimizing edge bending, or “soft clamping,” \textit{D}$_q$ can be significantly enhanced. This technique, initially demonstrated with phononic crystals (PnCs) featuring central defects \cite{bib15}, has been refined with strain engineering \cite{bib25, bib41} and mass engineering \cite{bib16}. Other methods, including fractal-like \cite{bib42} and hierarchical \cite{bib43} structures, as well as perimeter mode designs \cite{bib44}, have led to advanced resonator configurations that effectively minimize edge bending.

While analytical models provide insight into simple geometries, they often fall short in accounting for the complex geometry of advanced resonator designs, particularly those using phononic crystals. For such structures, numerical simulations are essential. Various optimization strategies have been employed to enhance resonator performance, including tuning the sizes of PnC cells and defects \cite{bib25}, experimenting with different PnC configurations \cite{bib15}, refining clamping geometries \cite{bib46}, and systematically adjusting parameters across the resonator \cite{bib13}. These approaches, often starting from heuristic design estimates and sometimes enhanced with machine learning \cite{bib47}, have yielded resonators with exceptional \textit{Q} and \textit{Qf} products.

Despite these achievements, unexplored designs with potentially superior performance may remain undiscovered. This limitation arises from two main issues. First, the initial predefined geometries typically constrain the optimization to a few variables, such as width, length or radius of curvature, restricting the scope of potential design space. Second, the gradients of \textit{Q} or \textit{Qf} with respect to these variables are often difficult to calculate, making many well-established gradient-based optimization algorithms unsuitable for this problem. 

These challenges can be tackled using methods like topology optimization \cite{bib4}, a numerical approach that seeks configurations with optimized performances without requiring a predefined geometry. It allows the topology of the structure to evolve throughout the optimization process, accommodating significant design changes. Moreover, topology optimization can be solved with various gradient-based algorithms. Recent studies \cite{bib7, bib11} have applied this method, using finite element analysis (FEA) \cite{bib20,bib21} to evaluate \textit{Q} at the fundamental eigenmode as the optimization objective. The resulting optimized designs exhibit features similar to trampoline resonators \cite{bib13} near the central pad, validating the design approach, while also introducing novel structural details near the outer boundary. Compared to similar dimensioned trampolines \cite{bib13} these optimized structures achieve significantly improved \textit{Qf} products, enabling the resonators to meet the demanding requirement of $Qf>6\times10^{12}$Hz at room temperature \cite{bib7, bib11}.

Most studies focus on the fundamental mode \cite{bib13,bib7,bib11}, or the fundamental mode within a band gap created by a phononic crystal \cite{bib25,bib16,bib39}. Higher-order eigenmodes are rarely explored and typically evaluated in only a few studies \cite{bib25,bib16,bib39}. 
The focus on the fundamental mode is often due to its higher effective mass, making it easier to excite, and its straightforward identification during eigenfrequency analysis, as it avoids the complex behavior associated with higher modes, such as mode interactions \cite{bib49}.
%On one hand, fundamental mode often acquires the highest effective mass and thus the easiest to stimulate. On the other hand, fundamental mode is usually convenient to identify in eigenfrequency analysis without sophisticated behaviours of higher eigen modes such as mode interactions \cite{bib49}.
However, exploring higher-order modes can be valuable. One reason is that achieving the required \textit{Qf} threshold to enter the quantum regime at room temperature necessitates balancing \textit{Q} and \textit{f}, a trade-off that topology optimization can address. Additionally, higher-order modes exhibit more intricate mode profiles than the fundamental mode, potentially reducing edge bending near clamping boundary without needing a photonic crystal structure. This could lead to more compact resonator designs. 

In this study, the resonator is designed for application in optical cavities, potentially using a membrane-in-the-middle configuration. Therefore, the resonator is two-dimensional (2D) with a central pad, enabling coupling with a laser through radiation pressure forces \cite{bib12}. Only intrinsic damping, primarily due to surface losses \cite{bib27}, is considered and is quantified by the dilution factor \textit{D}$_q$. External losses, such as gas damping, are mitigated by placing the resonator in a high-quality vacuum environment. In the finite element analysis (FEA), edge bending caused by high pre-stress \cite{bib2} is taken into account, and a locally refined mesh is used to capture this effect accurately. Additionally, we explore the topology optimization of resonator targeting higher-order eigenmodes, with \textit{D}$_q$ as the optimization objective. 

\section{Results}\label{sec2}
\subsection{Numerical simulation and optimization}\label{subsec1}
\begin{figure}[h]
\centering
\includegraphics[width=0.95\textwidth]{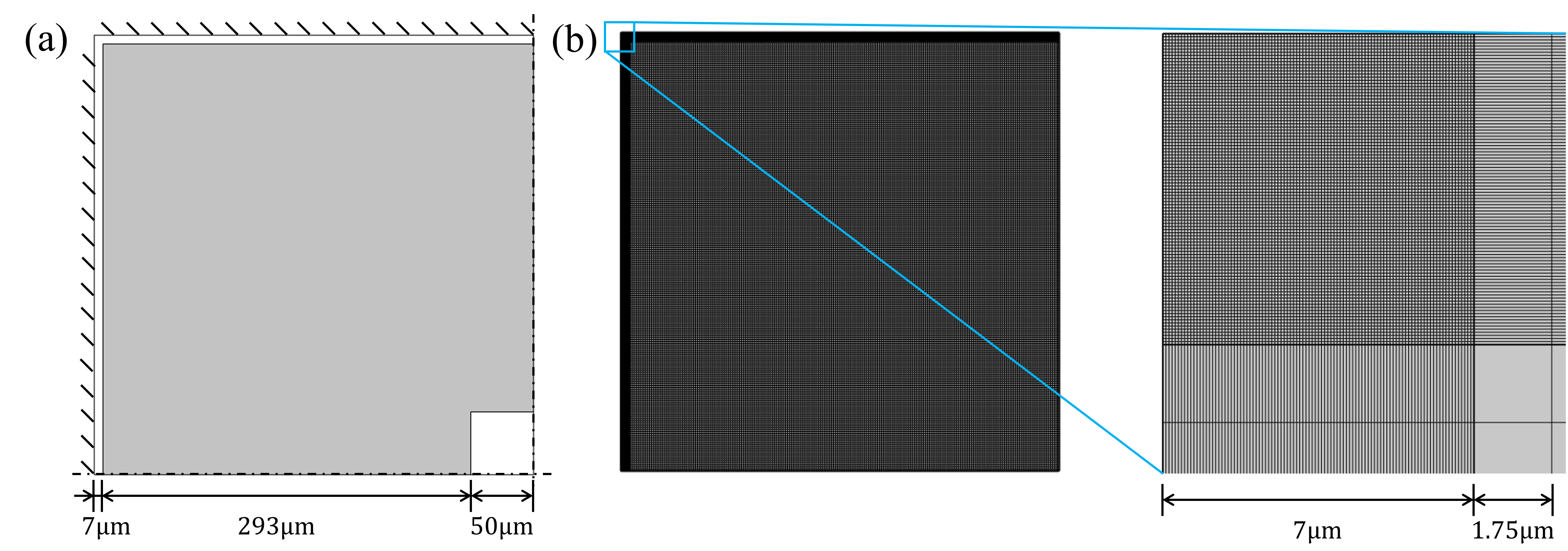}
\caption{Specifications of numerical model. (a) Dimensions and boundary conditions: The left and upper boundaries are clamped while the other two are set as symmetric conditions. The gray part represents the design domain, while the white regions indicate the non-design domain. 
(b) Mesh details: The complete mesh (left) and the locally refined mesh near the corner point (right). The locally defined region spans 7$\upmu$m with 100 elements, each with a resolution of 70nm. The coarser mesh has a resolution of 1.75$\upmu$m, resulting in a total mesh size of 296$\times$296 quadrilateral elements.}\label{fig1}
\end{figure}

The numerical model used for finite element analysis and topology optimization is detailed in Fig. \ref{fig1}. The resonator is composed of a pre-stressed Si$_3$N$_4$ layer with a nominal thickness of 50nm, a Young's modulus of 250GPa, a material density of 3100kg/m$^3$, and a Poisson's ratio of 0.23. The intrinsic quality factor $Q_0$ is set to 4000 based on reported results \cite{bib7}. To obtain compact device design, the resonator is embedded within a square window of 700$\times$700$\upmu$m$^2$. A 7$\upmu$m rim is included to accommodate variations in window size due to changes in silicon wafer thickness, and a central pad of 100$\times$100$\upmu$m$^2$ facilitates coupling between light and mechanical resonator via radiation pressure force \cite{bib12}. The rim and central pad remain fixed during topology optimization while the remaining area serves as the design domain where the design variables are iteratively updated. To optimize computational efficiency, only a quarter of the domain is analyzed, applying the boundary conditions shown in Fig. \ref{fig1}(a). Due to the high pre-stress from Si$_3$N$_4$ deposition, the curvature of mode shape changes abruptly near the fixed boundary \cite{bib2} and thus a locally refined mesh is applied on the rim while a coarser mesh is used for the rest of the domain, as shown in Fig. \ref{fig1}(b).  
\begin{figure}[h]
\centering
\includegraphics[width=1.0\textwidth]{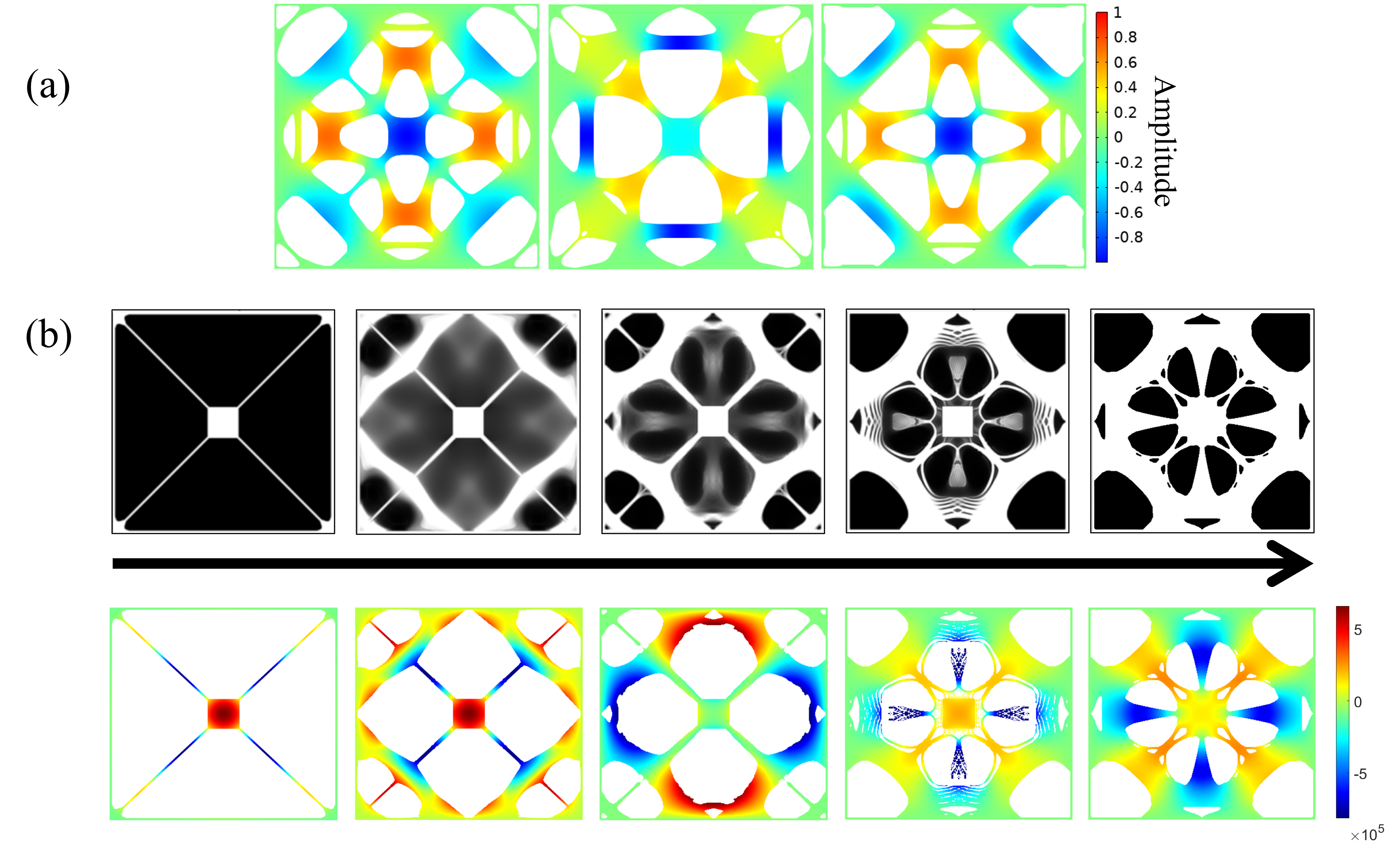}
\caption{Configurations of optimized resonators. (a) Design 2, 3 and 4, showing the normalized targeted eigenmode profiles of out-of-plane displacement plotted on the smoothed optimized designs. (b) Upper: evolution of design 1 from the initial guess to the final optimized design. Black indicates voids, while white represents the Si$_3$N$_4$ layer. In the final optimized structure, small holes were manually filled to facilitate fabrication. Lower: corresponding targeted eigenmodes plotted on elements with physical design variable $\bar{\rho}_e>0.65$. }\label{fig2}
\end{figure}

The intrinsic loss is the only loss mechanism considered in this study, modeled as being proportional to linear strain energy at the targeted eigenmode. To enable effective opto-mechanical coupling through radiation pressure forces, the central pad must maintain a nonzero out-of-plane displacement, as shown in Fig \ref{fig2}a. In some cases, such as in designs 1 and 3, the out-of-plane motion of the central pad does not reach the maximum magnitude, but this does not pose a problem for ringdown measurements and subsequent applications. The overall quality factor is estimated as $Q=Q_0\times D_q$ where $Q_0$ is the intrinsic quality factor and $D_q$ is the damping dilution factor. The material distribution is optimized by maximizing \textit{$D_q$} via density-based topology optimization, with a maximum material occupation ratio set to 50$\%$. The frequency of the fundamental mode is constrained to ensure structural connectivity. Details of topology optimization are provided in the Methods section. Different designs are achieved using various initial guesses: Design 1 starts from the nominal design reported by Norte et al.\cite{bib13}; Design 2 begins with a homogeneous square membrane. Design 3 uses the same initial guess as Design 1 but is optimized for a higher targeted eigenmode, while Design 4 starts with the same initial guess as Design 2 but with a modified feature selection methods that removes small tethers at the corners and near the central pad. The optimization evolution of Design 1 is shown as an example in Fig. \ref{fig2}(b), illustrating the progression from the initial guess to the final optimized configuration. 

\subsection{Fabrication and characterization}\label{subsec2}
\begin{figure}[H]
\centering
\includegraphics[width=1\textwidth]{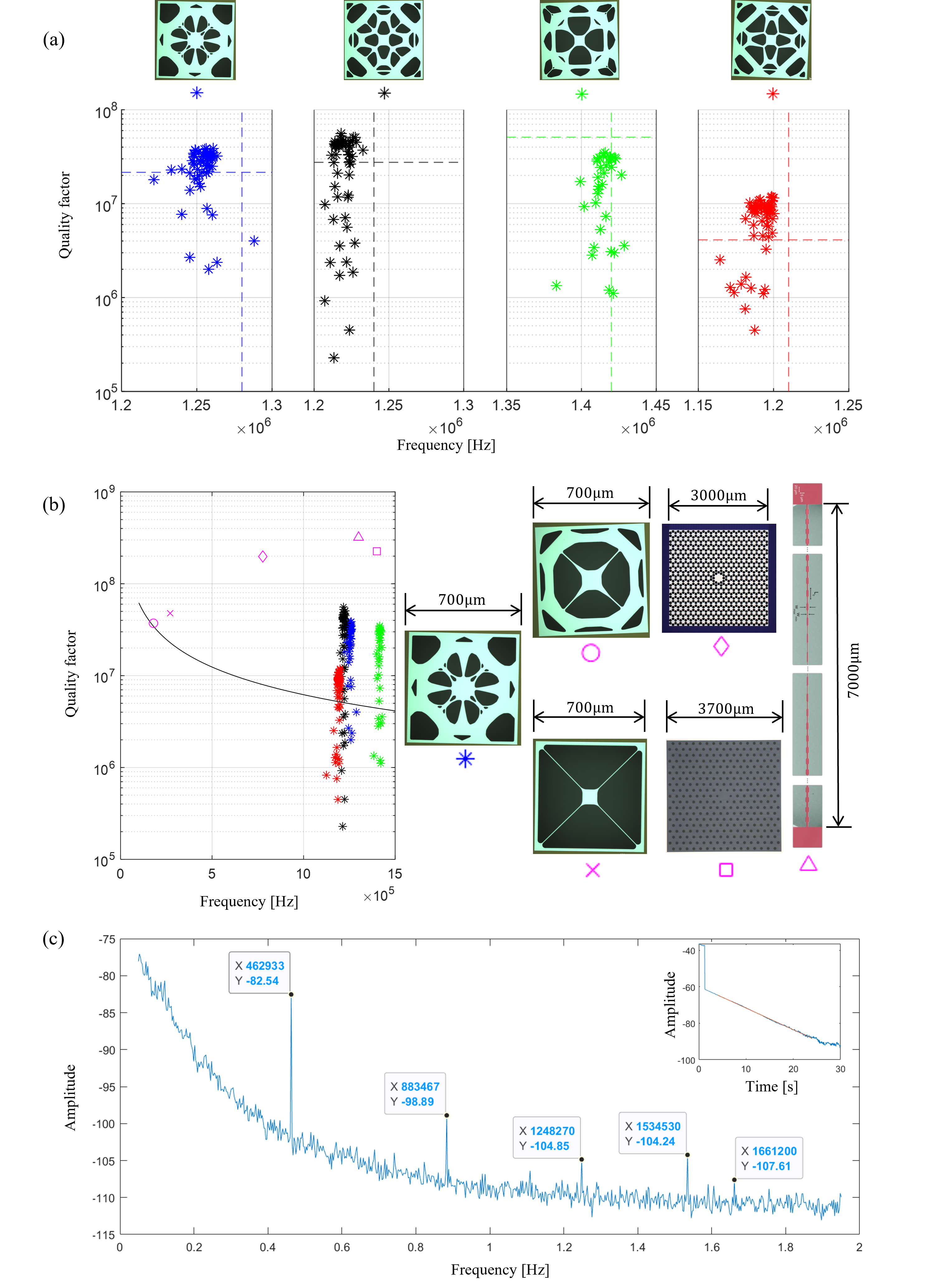}
\caption{(a) Ringdown measured \textit{Q} and \textit{f} values. Optical microscope images for design 1-4 are shown on top, where the white regions represent the Si$_3$N$_4$ layer, and black regions indicate voids. Data points correspond to measured values, while the dashed lines represent the numerically calculated \textit{Q} (horizontal) and \textit{f} (vertical) using a body-fitted mesh in COMSOL Multiphysics, as listed in Table \ref{table1}. (b) Comparison with reported results: \textit{Qf} (left) and size (right). The black solid lines indicates the contour where \textit{Qf}$=$6.2$\times$10$^{12}$. The yield represents the number of samples with \textit{Qf} exceeding this limit out of the total measured samples for each design: 54/58 for Design 1, 48/57 for Design 2, 27/35 for Design 3, and 46/62 for Design 4. The highest measured \textit{Qf} values of the designs are highlighted with magenta circle \cite{bib11} and crosses \cite{bib13}. \textit{Q} values of other designs, indicated by magenta diamond \cite{bib15}, squares \cite{bib16} and triangles \cite{bib25} are scaled to thickness of 50nm for consistency with design 1-4. (c) A representative power spectrum of Design 1. The inset shows the corresponding ringdown curve over 30s at the targeted mode, with \textit{f}: 1.248MHz and \textit{Q}: 29.13$\times10^{6}$.}
\label{fig3} 
\end{figure}

The optimized structures were patterned on a 50nm Si$_3$N$_4$ layer deposited by low pressure chemical vapor deposition (LPCVD) with a pre-stress of approximately 1.2GPa. Both the pre-stress and thickness exhibited minimal variation due to fluctuations in the deposition process. For comparison, two additional resonator designs from previous studies \cite{bib11, bib13} were fabricated alongside the optimized structures, as shown in Fig. \ref{fig3}(b). 

Ringdown measurements were conducted in a vacuum system with a pressure below $10^{-6}$ mbar at room temperature. The measured \textit{Q} and \textit{f} values are presented in Fig. \ref{fig3}(a). For designs 1-4, more than 75$\%$ of the measured samples met the minimum requirement for entering the quantum regime at room temperature, defined by \textit{Qf}$>6.2\times10^{12}$ \cite{bib12}. A power spectrum and the corresponding ringdown curve for Design 1 are shown in Fig. \ref{fig3}(c). The measured \textit{Q} values display some variation, likely due to contamination by microscopic particles, which can result from cross-contamination during wet-chemical processing or exposure during storage. Given that the quality factor due to gas damping is proportional to the operating frequency of the resonator, higher pressure conditions can be tolerated compared to those reported in previous studies \cite{bib11}. With a pressure below $10^{-6}$ mbar, gas damping is effectively mitigated and, therefore, not included in the numerical model. Another potential loss mechanism is phonon tunneling loss (PTL) \cite{bib2}, which arises from energy radiation from the resonator to the substrate. However, this effect was also excluded from the numerical model. The close agreement between the \textit{Q} values measured through ringdown and those predicted by the numerical model indicates that PTL is not a dominant source of loss and can be safely neglected. As shown in Fig. \ref{fig3}(a), designs 1,2, and 4 exhibit a slight discrepancy between the numerically estimated \textit{Q} values and those measured through ringdown, where the former should represent an upper bound due to the exclusion of PTL. The numerical estimates assume $Q = Q_0\times D_q$, with $Q_0 = 4000$ based on previously reported values \cite{bib7,bib11,bib13}. In reality, $Q_0$ can vary between 3000 to 7000 across different devices. When the upper bound of $Q_0$ is considered, the discrepancy between the numerical and experimental results is resolved. 

Comparisons of \textit{Qf} product and size between the proposed designs and previously reported results are shown in Fig. \ref{fig3}(b). The proposed designs demonstrate an increased \textit{Qf} product compared to resonators with similar dimensions. This improvement is primarily due to an increase in the eigenfrequency, while the \textit{Q} factor is either optimized (design 1, 2 and 4) or maintained at a similar level (design 3). In contrast, designs from \cite{bib15,bib16,bib25} exhibit higher \textit{Q} and \textit{Qf} values than those of the proposed designs. This advantage stems from the use of phononic crystals, which effectively eliminate edge-bending losses near the clamping boundaries, a limitation that still affects the proposed designs. Additionally, a larger aspect ratio \cite{bib25} contributes to a higher \textit{Q} factor. Despite these differences, the proposed designs may still be preferred in some scenarios where compact devices are required. Furthermore, in nano-fabrication, a large aspect ratio significantly increases the complexity and reduces the yield, which could be a critical drawback for large-scale production or practical applications. Thus, the proposed designs offer a balance between performance and manufacturability, making them advantageous in specific use cases. 

\section{Discussions}\label{sec3}
Several parameters and figures of merit are numerically evaluated in Table \ref{table1} using COMSOL Multiphysics, with the simulation geometry precisely matching that used in fabrication. The proposed resonators, characterized by high \textit{Q} and \textit{f}, offer significant advantages for applications in quantum opto-mechanics and sensing. Operating at higher-order modes rather than the fundamental mode results in substantial reduction in the effective mass at targeted mode, making this approach more effective than simply reducing the total mass. However, this benefit comes with the price of increased frequency. For the most optimal design among the proposed resonators (Design 2), the requirements for ground-state cooling and force sensitivity are improved nearly fivefold compared to Trampoline 2. Specifically, the ground-state cooling requirement, $\sqrt{S^{imp,gs}_{xx}} = \sqrt{2 \hbar^2 Q_0 /(k_B T_0 \omega m_{eff})}$, and the force sensitivity, $\sqrt{S^{th}_{FF}} = \sqrt{4 k_B T_0 m_{eff}\omega/Q}$, are both significantly enhanced, where $\omega = 2\pi f$, $T_0=300$K, $k_B$ is Boltzmann’s constant, and $\hbar$ is the reduced Planck constant.

\begin{table}[h]
\caption{Parameters and figures of merit}
\begin{tabular}{@{}llllll@{}}
\toprule
  & \textit{f} [MHz] & \textit{Q} ($\times$10$^6$) & \textit{$m_{eff}$} [kg] & $\sqrt{S^{imp,gs}_{xx}}$ [$\mbox{m}/\sqrt{\mbox{Hz}}$] & $\sqrt{S^{th}_{FF}}$ [$\mbox{N}/\sqrt{\mbox{Hz}}$] \\
\midrule
Design 1                       & 1.28 & 21.55 & 7.35$\times$$10^{-14}$ & 1.40$\times$$10^{-17}$ & 2.13$\times$$10^{-17}$   \\
Design 2                       & 1.24 & 27.53 & 3.21$\times$$10^{-14}$ & 2.43$\times$$10^{-17}$ & 1.23$\times$$10^{-17}$   \\
Design 3                       & 1.42 & 51.02 & 5.39$\times$$10^{-14}$ & 2.38$\times$$10^{-17}$ & 1.25$\times$$10^{-17}$   \\
Design 4                       & 1.21 &  4.12 & 1.58$\times$$10^{-13}$ & 4.28$\times$$10^{-18}$ & 6.96$\times$$10^{-17}$   \\
Trampoline 1 \cite{bib11}      & 0.28 & 29.22 & 7.10$\times$$10^{-12}$ & 3.53$\times$$10^{-18}$ & 8.44$\times$$10^{-17}$   \\
Trampoline 2 \cite{bib13}      & 0.19 & 19.77 & 2.51$\times$$10^{-12}$ & 6.00$\times$$10^{-18}$ & 4.97$\times$$10^{-17}$   \\
\botrule
\end{tabular}
\label{table1}
\end{table}

Previously, high \textit{Qf} resonators \cite{bib7,bib11} were obtained via topology optimization at fundamental frequencies below 500kHz. Higher-frequency resonators were typically realized using phononic crystal methods combined with heuristic optimizations \cite{bib15,bib16}, which resulted in larger dimensions ($\sim$mm) compared to those obtained through topology optimization ($700\upmu$m). The approach proposed in this article combines the strengths of both methods, allowing for the specification of any eigenmode, provided that its mode profile supports effective opto-mechanical coupling, while simultaneously optimizing \textit{Q} at the targeted mode. The resulting resonators feature compact dimensions, high eigenfrequencies, and high \textit{Q} factors. The power spectrum shown in Fig. \ref{fig3}(c) demonstrates that the targeted mode is well isolated from neighbour modes, effectively replicating the band gap behavior achieved with phononic crystal designs. This method also has the potential to extend the targeted mode to even higher frequencies, such as 10MHz, by carefully selecting an appropriate eigenmode profile. 

The proposed FEA and topology optimization approach is highly adaptable to various extensions. Due to the nature of the opto-mechanical coupling, the targeted mode always exhibits out-of-plane displacement on the central pad. In other applications, such as ultra-fast force microscopy \cite{bib18} or attonewton-scale force sensitivity measurements \cite{bib19}, a torsional rotation may be preferred \cite{bib17}. This can be achieved by defining a torsional mode as the targeted in the optimization process. Additionally, other parameters, such as $\sqrt{S^{imp,gs}_{xx}}$ and $\sqrt{S^{th}_{FF}}$ can be incorporated as constraints in the optimization model to tailor performance further. 

Currently, the optimization focuses solely on the geometry of the resonator. However, other factors, such as variations in material properties or crystal orientations during fabrication, could be included by introducing additional sets of design variables. Furthermore, this model could be integrated with phononic crystal methods, for example, by optimizing the geometry or distribution of phononic crystal cells operating at different frequencies, or by refining the geometry of defects within the crystal structure. This hybrid approach could further enhance the performance and versatility of the resonator designs.  

\section{Methods}\label{sec4}
\subsection{Finite element analysis}\label{subsec3}
Finite element analysis (FEA) is used to solve the pre-stressed resonator model. To prevent shear-locking in the extremely thin membrane, we employ quadrilateral MITC4 (Mixed Interpolation of Tensorial Components) cell elements \cite{bib1}. Moreover, to suppress instabilities in the eigenfrequency analysis, the translational displacements are interpolated using quadratic shape functions, while the rotational displacements are interpolated using linear shape functions. The high pre-stress resulting from the Si$_3$N$_4$ layer deposition causes abrupt changes in the curvature of the eigenmode profile near the fixed boundaries \cite{bib2}. To accurately capture these localized features, a locally refined mesh is employed in the regions near the fixed boundaries. 

The FEA process is composed of two steps. The first step is a linear static analysis to determine the distribution of pre-stress once the resonator is released. The second step is an eigenvalue analysis that incorporates the pre-stress \cite{bib7,bib29}, solving for the targeted eigenfrequency and the corresponding eigenmode profile: 
\begin{equation}
\boldsymbol{K}_0\boldsymbol{U}_0=\boldsymbol{F}_0 (\boldsymbol{\sigma}_0)
\label{eq1}
\end{equation}
\begin{equation}
\left(\boldsymbol{K}_0 + \boldsymbol{K}_\sigma(\boldsymbol{U}_0) - \omega_j^2 \boldsymbol{M} \right) \boldsymbol{\phi}_j  = \boldsymbol{0} 
\label{eq2}
\end{equation}
Here, $\boldsymbol{K}_0$ is the linear system stiffness matrix, $\boldsymbol{F}_0$ is the load vector derived from the initial stress, $\boldsymbol{K}_\sigma$ is the stress stiffness matrix dependent on the initial stress $\boldsymbol{\sigma}_0$ and the displacement vector $\boldsymbol{U}_0$, and $\boldsymbol{M}$ represents the consistent mass matrix of the system. The terms $\omega_j$ and $\boldsymbol{\phi}_j$ correspond to the eigenfrequency and eigenmode profile of the \textit{j}th mode, which is the targeted mode. The quality factor, \textit{Q}, is defined as $\textit{Q} = \textit{Q}_0*D_q$, where $\textit{Q}_0 = 4000$ is the intrinsic quality factor and $D_q$ is the damping dilution factor \cite{bib2}, calculated at the \textit{j}th eigenmode as:
\begin{equation}
D_q = \frac{\boldsymbol{\phi}_j^T \boldsymbol{K}_\sigma \boldsymbol{\phi}_j}{\boldsymbol{\phi}_j^T \boldsymbol{K}_0 \boldsymbol{\phi}_j} = \frac{\omega_j^2\boldsymbol{\phi}_j^T \boldsymbol{M} \boldsymbol{\phi}_j}{\boldsymbol{\phi}_j^T \boldsymbol{K}_0 \boldsymbol{\phi}_j} -1
\label{eq3}
\end{equation}
This formulation accounts for the influence of the pre-stress on the resonator's stiffness and the subsequent effect on the targeted eigenmode.

\subsection{Topology optimization}\label{subsec4}
The method of topology optimization employed here is density-based\cite{bib5}, which transforms the binary 0-1 material distribution problem into a continuous optimization problem. The three-field topology optimization scheme \cite{bib3} defines a design variable $\rho_e\in[0,1]$ for each element in the FEA mesh, representing the material occupancy of that element. To avoid checkerboard pattern and ensure mesh dependence\cite{bib4}, the design variables are first filtered and then projected to improve discreteness as follows: 
\begin{equation}
\label{eq4}
\tilde{\rho}_e = \frac{\sum_{i\in{n_e}}(\rho_{i}v_{i}H_{e,i})}{\sum_{i\in{n_e}}(v_{i}H_{e,i})}, 
\bar{\rho}_e=\frac{\tanh(\beta\eta)-\tanh\left(\beta(\tilde{\rho}_e-\eta)\right)}{\tanh(\beta\eta)-\tanh\left(\beta(1-\eta)\right)}
\end{equation}
where $n_e$ is the set of neighboring elements whose center-to-center distance $\Delta(e,i)$ is smaller than the filter radius $r_{min}=5\upmu m$. $H_{e,i}$ is the weight function defined as $H_{e,i}=\max{(0,r_{min}-\Delta(e,i))}$, and $v_i$ represents the volume of element $i$. The parameter $\beta$ controls the sharpness of smoothed Heaviside function and is updated during the optimization process, while $\eta$ serves as the threshold. $\bar{\rho}_e$ is the physical density variable used for material distribution. As $\beta$ increases during the optimization loop, the filtered design variables $\tilde{\rho}_e>\eta$ are interpreted as Si$_3$N$_4$ layers with $\bar{\rho}_e\approx1$, while $\tilde{\rho}_e<\eta$ are interpreted as voids with $\bar{\rho}_e\approx0$. 

The Young's modulus is interpolated using the Rational Approximation of Material Properties (RAMP) method \cite{bib6}, while the material mass density is linearly interpolated: 
\begin{equation}
\begin{split}
E_e &= \frac{\bar{\rho}_e}{1+q(1-\bar{\rho}_e)}(E_0 - E_{\min})+E_{\min},     q = 3
\\
\varrho_e &= \varrho_{\min} + (\varrho_0-\varrho_{\min})\bar{\rho}_e, 
\bar{\rho}_e \in [0,1]
\end{split}
\end{equation}
where $E_{\min} = 10^{-6}E_0$ and $\varrho_{\min} = 10^{-7}\varrho_0$ are used to suppress spurious modes in the eigenfrequency analysis caused by inappropriate stiffness-to-mass ratios in low-density regions. To reduce wrinkling-like instabilities in low-density regions, we apply displacement field interpolation as described in Gao et al. \cite{bib7}. Instabilities in high-density regions are controlled by employing mixed formulation elements, where the transnational degrees of freedom (DoFs) are interpolated with quadratic shape functions, while the rotational DoFs are interpolated using linear shape functions. This approach ensures accurate modeling across varying density regions, maintaining stability throughout the optimization process. 

To impose length scale constraints on the blueprint design and ensure manufacturability -- particularly limited by the resolution of photoresist exposure -- a geometry constraint \cite{bib8} is applied. The optimization problem is formulated as follows: 
\begin{eqnarray*}
&\underset{\rho_e}\max: {D_q(\boldsymbol{\phi}_j)} \\
&s.t.: 
\boldsymbol{K}_0(\bar{\rho}_e)\boldsymbol{U}_0=\boldsymbol{F}_0(\bar{\rho}_e), \\
&\left(\boldsymbol{K}_0 + \boldsymbol{K}_\sigma(\bar{\rho}_e,\boldsymbol{U}_0) - \omega_j^2 \boldsymbol{M}(\bar{\rho}_e) \right) \boldsymbol{\phi}_j  = \boldsymbol{0}, 
\\
&\omega_j\geq\bar{\omega}_j, \\
&\omega_{j-1}\leq(1-\varepsilon)\omega_j, \omega_{j+1}\geq(1+\varepsilon)\omega_j, \\
&\omega_{1}\geq\bar{\omega}_1, \\
&g_s\leq\epsilon, g_v\leq\epsilon, \\
&\frac{\sum_{e}(\bar{{\rho}}_{e}v_{e})}{\sum_{e}v_{e}}\leq\bar{V}, \\
& 0\leq{\rho}_e\leq1
\end{eqnarray*}
where $\bar{\omega}_j$ and $\bar{\omega}_1$ represent the lower frequency limit for the targeted and fundamental modes, respectively; $\bar{V}$ is the upper limit of volume fraction; $g_s$ and $g_v$ are geometry constraint functions \cite{bib8} for the solid and void phases, respectively; $\epsilon = 10^{-5}$ is a small number used to account for numerical errors; and $\varepsilon = 0.005$ is a small positive value that ensures separation between neighboring modes and the targeted mode. This formulation ensures that the design is not only optimized for the desired eigenmode but also remains feasible for fabrication, with appropriate control over dimensions and material distribution. 

The geometry constraint is introduced when $\beta$ is large, and the topology of the design has become relatively well-defined. However, even with a large $\beta$, topology optimization may still result in some elements with intermediate design variables. To address this, a double filter method \cite{bib30} is used prior to applying the geometry constraint. By adjusting the projection threshold defined in Eq.\ref{eq4}, intermediate elements can be selectively retained or removed, allowing for the fine-tuning of design features. The targeted mode is initially selected according to the eigenmode $\phi_j$ of initial guess, ensuring that the displacement on the central pad remains out-of-plane. As the topology evolves during optimization, $\phi_j$ also changes accordingly. To maintain continuity in mode identification, a mode tracing method \cite{bib14} is applied to track the eigenmode profile that most closely matches the one from the previous optimization step. The targeted mode is separated from neighboring modes because, in opto-mechanical applications, it is preferable for the mode to be distinct; this makes it easier to excite and helps maintain the desired $\phi_j$. The multi-constraint optimization problem is solved using the method of moving asymptotes (MMA) \cite{bib10}, which relies on the gradient information of the objective function and constraints derived using the adjoint method \cite{bib22} and chain rules \cite{bib31}. Due to the FEA discretization, stair-step features may appear at the boundaries between solid and void regions, potentially leading to local stress concentrations. To mitigate this, the blueprint designs are smoothed by removing convex features and applying a low-pass filter, following techniques similar to those described in Høj et al. \cite{bib11}. This approach ensures smoother final design while preserving the structural integrity and performance of the resonator.

\subsection{Fabrication and characterization}\label{subsec5}
The fabrication process starts by depositing a 50nm thick Si$_3$N$_4$ layer as the resonator material on a 4-inch silicon wafer, 500$\upmu$m thick, using low-pressure chemical vapor deposition (LPCVD). A layer of photoresist is deposited on top of the Si$_3$N$_4$, followed by selective exposure and development using tetramethylammonium hydroxide (TMAH). The exposed areas are etched away using reactive ion etching (RIE) to define the resonator structure, and the remaining photoresist is removed with oxygen plasma. The wafer is then diced into individual chips, which are mounted onto custom-designed holders. The resonators are released from the silicon substrate by etching in potassium hydroxide (KOH) at 60$^{\circ}$C, followed by cleaning with hydrochloric acid and a solution of Piranha. The samples are dried using ethanol vapor for 20 minutes to prevent stiction. 

For characterization, a 1550nm laser is directed onto the vibrating resonator, and the resulting phase shift due to the motion is detected with high sensitivity using a phase-locked homodyne detector. The resonator is driven at its resonance frequency, and once resonance is achieved, the excitation is abruptly switched off. The decay of the resonator's amplitude is recorded, allowing for the evaluation of the the quality factor \textit{Q}. 

% \bibliography{sn-bibliography}
%% BioMed_Central_Bib_Style_v1.01

\section{Acknowledgements}\label{sec5}
The authors would like to thank Wenjun Gao for his numerical implementations in MATLAB. 

\section{Conflict of Interest}\label{sec6}
The authors declare no conflict of interest.

\end{document}